\documentclass[twocolumn]{aastex6}
\usepackage{apjfonts,graphicx}

\newcommand{\mum}{$\mu$m}

\newcommand{\muasyr}{$\mu$as yr$^{-1}$}

\newcommand{\kms}{km~s$^{-1}$}

\shorttitle{Water Masers, Ammonia, and Hydrogen Recombination Lines in M31}
\shortauthors{Darling, Amiri, Gerard, \& Lawrence}

\begin{document}

\title{Water Masers in the Andromeda Galaxy:  
I.  A Survey for Water Masers, Ammonia, and Hydrogen Recombination Lines} 

\author{Jeremy Darling, Benjamin Gerard\altaffilmark{1}, 
Nikta Amiri\altaffilmark{2}, 
\& Kelsey Lawrence}
\affil{Center for Astrophysics and Space Astronomy,
Department of Astrophysical and Planetary Sciences,
University of Colorado, 389 UCB, Boulder, CO 80309-0389, USA}
\email{jdarling@colorado.edu}
\altaffiltext{1}{Current address:  University of Victoria, Department of Physics and Astronomy, 3800 Finnerty Rd., Victoria, V8P 5C2, Canada}
\altaffiliation[\altaffilmark{2}]{Current address:  Jet Propulsion Laboratory, M/S 238-600, 4800 Oak Grove Dr., Pasadena, CA 91109, USA}

\begin{abstract}
We report the results of a Green Bank Telescope survey for water masers, 
ammonia (1,1) and (2,2), and the H66$\alpha$ recombination line toward 
506 luminous compact 24~\mum-emitting regions in the Andromeda Galaxy (M31).  
We include the 206 sources observed in the Darling (2011) water maser survey for completeness.
The survey was sensitive enough to detect any maser useful for $\sim$10~\muasyr\
astrometry.  
No new water masers, ammonia lines, or H66$\alpha$ recombination lines were detected individually
or in spectral stacks reaching rms noise levels of $\sim$3 mJy and $\sim$0.2 mJy, respectively, in 
3.1--3.3 km s$^{-1}$ channels.  The lack of detections in individual spectra and in the spectral stacks
is consistent with Galactic extrapolations. 
Contrary to previous assertions, there do not seem to be additional bright water masers to be found 
in M31.  The strong variability of water masers may enable new maser detections in the future, but variability
may also limit the astrometric utility of known (or future) masers since flaring masers must also fade. 
\end{abstract}


\section{Introduction}

The dominant galaxies in the Local Group, 
the Milky Way and the Andromeda galaxy (M31), are likely to merge
in 5.86$^{+1.61}_{-0.72}$ Gyr \citep{vandermarel12c}.  The key
quantity for this prediction is the transverse velocity of M31 with respect to 
the Galaxy.
\citet{sohn12} and \citet{vandermarel12b} used {\it Hubble Space
Telescope} observations of thousands of M31 stars referenced to hundreds of compact 
background galaxies in three fields (near the major and minor axes and in the Giant Southern Stream) 
over five to seven years 
to constrain the tangential velocity of M31 to $\leq34.3$~km~s$^{-1}$ ($1\sigma$).  
The consistency of the M31-Milky Way velocity with a radial orbit,
infalling at $-109.3\pm4.4$~km~s$^{-1}$ \citep{vandermarel12b}, 
provides a new estimate of the Local Group timing mass 
to be $(4.93\pm1.63)\times10^{12}\ M_\odot$ \citep{vandermarel12b}, 
although \citet{vandermarel12b} derive a best estimate of $(3.17\pm0.57)\times10^{12}\ M_\odot$,
citing cosmic variance as an uncertainty floor on the timing argument estimation of the Local Group mass.
The transverse velocity of M31 is thus essential to our understanding of 
the Local Group's future (and past) dynamical evolution as well as the density 
profiles and distribution of dark matter halos \citep{peebles01,loeb05,reid09}. 

The \citet{sohn12} and \citet{vandermarel12b,vandermarel12c} results can be independently verified and possibly refined using 
maser proper motions in M31.  Molecular masers, while small in number, are compact 
high brightness temperature light sources arising in the molecular ring of M31 
that provide $\sim$10~microarcsecond ($\mu$as) astrometry using Very Long Baseline 
Interferometry (VLBI).  \citet{brunthaler05,brunthaler07} have measured the proper motions of 
Galactic analog 22~GHz water (H$_2$O) masers discovered in M31's
satellite galaxies M33 \citep[first by][]{churchwell77} and IC~10 \citep{henkel86}, obtaining 
their 3-dimensional velocities with respect to the Milky Way.  
\citet{brunthaler05} also measured the proper rotation of 
M33, which provides a geometric determination of its distance via ``rotational parallax.''
The same can be done with masers in M31 \citep{darling11}.

Detecting Galactic analog masers of any type in M31 is challenging, due to its 
distance and large angular size.  The M31 distance reduces Galactic maser 
flux densities by a factor of $\sim$10$^4$, so only the brightest Galactic masers could
be detected in M31, provided that surveys are sensitive to $\sim$10 mJy spectral lines \citep{darling11}.  
The large angular size of M31 --- the molecular ring is roughly $2^\circ\times0\fdg5$ ---
implies that surveys for masers, particularly H$_2$O, need to be pointed in order to reach adequate 
sensitivity.  New radio telescope facilities with larger collecting areas or larger
fields of view than are currently available may improve this situation.

\begin{figure*}[ht]
\includegraphics[width=7in]{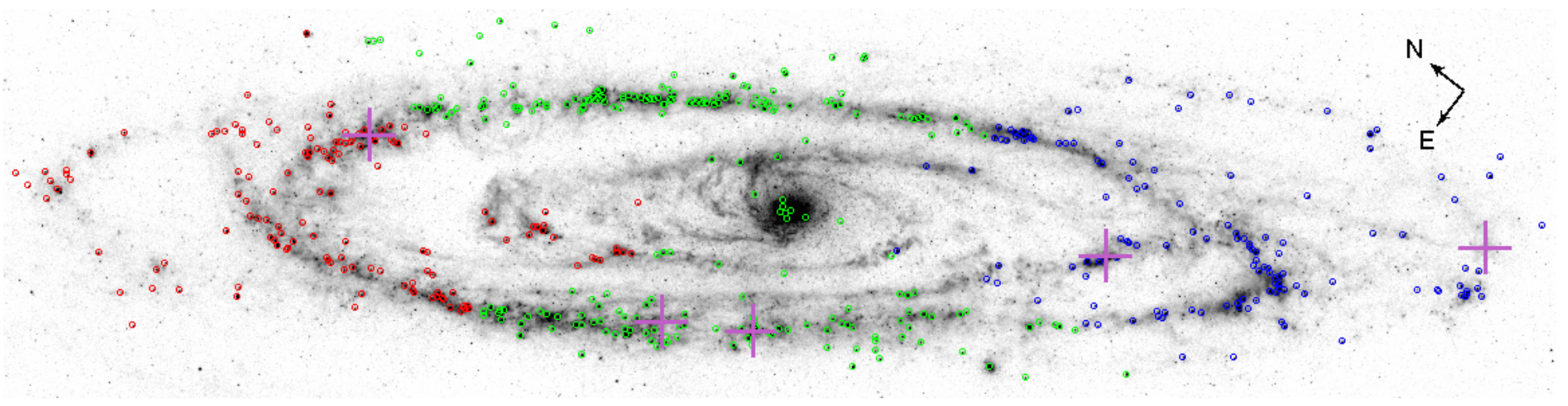}
\caption{
Spitzer 24~$\mu$m map of M31 \citep{gordon08}.
The circles indicate the 506 pointing centers 
(the circles are to scale, showing the 33\arcsec\ FWHM beam).  
Crosses mark the masers detected by \citet{darling11} and are enlarged for clarity.  
Colors indicate the systemic (green), red, and blue spectrometer
tuning centers at $-$300, $-$100, and $-$500~km~s$^{-1}$, respectively.
The image spans $2\fdg5\times 0\fdg6$.  
 }\label{fig:M31map}
\end{figure*}

\citet{sullivan73} may have made the first search for H$_2$O masers in M31, making no 
detections toward the nucleus and ``emission nebula \#132.''  
\citet{greenhill95} and \citet{imai01} report pointed 
H$_2$O maser surveys toward \ion{H}{2} regions in M31, reaching 1$\sigma$ rms noise levels of 
29 and 70~mJy, respectively.  
Claussen \& Beasley (private communication) surveyed the
nuclear region and most of the molecular ring using the Very Large Array, 
making no H$_2$O maser detections and reaching an rms noise of 30 mJy per beam.
\citet{sjouwerman10} detected the first maser of any type in M31, 
a Class II methanol (CH$_3$OH) maser at 6.7~GHz, but no associated water maser was detected
\citep[see also][and this work]{darling11}.
Finally, \citet{darling11} conducted a survey of 206 compact 24~$\mu$m sources in the molecular
ring with $\sim$3~mJy rms noise per 3.3 km s$^{-1}$ channel and detected five water masers.  
The water maser non-detections prior to 2011 were all consistent with 
the aggregate star formation rate of M31, and these surveys 
were not sufficiently sensitive to detect a typical Galactic maser 
at the distance of M31 \citep{brunthaler06}.  
The historical lack of Galactic analog water maser detections in M31 has simply
been a problem of sensitivity and of locating likely sites of maser activity, both of which
now have remedies.

In this paper,  we describe an extension of the \citet{darling11}
Green Bank Telescope\footnote{The National Radio Astronomy Observatory is a facility of the National
Science Foundation operated under cooperative agreement by Associated
Universities, Inc.} (GBT) H$_2$O maser survey
to include an additional 300 unresolved 24~$\mu$m sources in M31.    We also report on the simultaneous
observation of two ammonia (NH$_3$) lines and one hydrogen radio recombination line (H66$\alpha$)
toward all 506 locations in the combined surveys.  In a companion paper, 
we examine the physical conditions most likely to produce detectable 
H$_2$O maser emission in M31 \citep[][Paper \rm{II}]{amiri16}.

Throughout this manuscript, we use heliocentric velocities (optical definition)
and assume the following about M31:
a systemic velocity of $-300$~km~s$^{-1}$ \citep{devaucouleurs91},
a distance of 780~kpc \citep{mcconnachie05}, 
and central coordinates 00:42:44.3, +41:16:09 (J2000).

\section{Sample Selection}\label{sec:selection}
 
A Spitzer 24~$\mu$m map of M31 \citep{gordon08}
guided the GBT observations in two important ways:  source selection and completeness.  
The association of water masers with (ultra)compact \ion{H}{2} regions is well known, but
the molecular gas as traced by optically thick CO emission does not necessarily indicate these regions.
Likewise, H$\alpha$-selected \ion{H}{2} regions may select {\it against} 
dust-enshrouded ultra-compact \ion{H}{2} regions \citep[but see][]{amiri16}.
Compact 24~$\mu$m emission, however, is a good indicator of likely sites of 
H$_2$O maser emission, and there is a rough relationship between far-IR emission and 
water maser luminosity \citep[][although the latter relies on
bolometric luminosities]{jaffe81,urquhart11}.
We therefore selected compact 24~$\mu$m sources 
associated with the dusty molecular --- and presumably star-forming --- regions in M31
(``compact'' means unresolved at the 23.5~pc resolution of the Spitzer map).
This method can also, however, select objects unlikely to produce water masers, such as planetary nebulae, 
giant stars, and background galaxies.

We constructed a catalog of 506 unresolved 24~$\mu$m sources working from the brightest down to the point where most 
of the 24~$\mu$m emission becomes extended, about 4~MJy~steradian$^{-1}$ at peak specific 
intensity.
Although this pointed survey does not include all star formation in
M31, it does include the majority of likely locations for strong H$_2$O maser activity.  By working down 
in luminosity through the catalog of compact 24~$\mu$m sources, we have surveyed a large fraction of the 
total ongoing star formation in the galaxy \citep[e.g.,][]{tabatabaei10,ford13} and thus
 most of the likely H$_2$O-bearing regions \citep{amiri16}.   
Figure \ref{fig:M31map} shows the 24~$\mu$m map of M31 with the
506 pointing centers and primary beam size for our GBT water maser survey, including the masers
detected by \citet{darling11}.

\section{Observations and Data Reduction}\label{sec:obs}

\citet{darling11} observed 206 24~\mum\/ sources in M31 using the GBT in 2010 October through
December.  The $6_{16}-5_{23}$ 22.23508~GHz ortho-water maser line observations were reported 
in \citet{darling11}, but simultaneous observations of the 
para-ammonia (NH$_3$)  rotational ground state inversion transitions in the 
metastable states $(J,K) = (1,1)$ and $(2,2)$ at 23.6945 and 23.72263~GHz and the
hydrogen recombination line H66$\alpha$ at 22.36417~GHz were not.  
We subsequently observed all four of these lines toward an additional 300 
24~\mum\/ sources in 2011 October through 2012 January.  
The telescope configuration was identical to that described by \citet{darling11}:
A 632--674~km~s$^{-1}$ (50~MHz) bandpass was centered, for each of the four observed transitions,
on a heliocentric velocity of 
$-300$~km~s$^{-1}$ for 239 sources in the central parts of the galaxy and
along the minor axis, on 
$-100$~km~s$^{-1}$ for 136 sources in the redshifted northeast wedge of the galaxy, and on 
$-500$~km~s$^{-1}$ for 131 sources in the blueshifted southwest wedge of the galaxy 
(Figure \ref{fig:M31map} and Table \ref{tab:obs}).

\begin{deluxetable*}{lcccccc|lcccccc}
\tabletypesize{\scriptsize} 
\tablecaption{M31 Line Survey Results \label{tab:obs}}
\tablewidth{0pt}
\tablehead{
\colhead{Object} & \colhead{H$_{2}$O} & \colhead{NH$_{3}$} & \colhead{NH$_{3}$} & \colhead{H66$\alpha$}  & \colhead{$V_{\text{obs}}$} & \colhead{$V_{\text{CO}}$} &
\colhead{Object} & \colhead{H$_{2}$O} & \colhead{NH$_{3}$} & \colhead{NH$_{3}$} & \colhead{H66$\alpha$}  & \colhead{$V_{\text{obs}}$} & \colhead{$V_{\text{CO}}$} \\
\colhead{} & \colhead{} & \colhead{(1,1)} & \colhead{(2,2)} & \colhead{} & \colhead{} & \colhead{} &
\colhead{} & \colhead{} & \colhead{(1,1)} & \colhead{(2,2)} & \colhead{} & \colhead{} & \colhead{}\\
\colhead{(J2000)} & \colhead{(mJy)} & \colhead{(mJy)} & \colhead{(mJy)} & \colhead{(mJy)} & \colhead{(km s$^{-1}$)} & \colhead{(km s$^{-1}$)} & 
\colhead{(J2000)} & \colhead{(mJy)} & \colhead{(mJy)} & \colhead{(mJy)} & \colhead{(mJy)} & \colhead{(km s$^{-1}$)} & \colhead{(km s$^{-1}$)}}
\startdata 
003838.7+402613.5\tablenotemark{a} & 2.0 & 2.4 & 2.3 & 2.2 & $-$500 & \nodata & 
003849.2+402551.7\tablenotemark{a} & 3.2 & 2.7 & 2.6 & 3.2 & $-$500 & \nodata \\
003852.5+401904.9 & 2.0 & 2.1 & 2.4 & 2.0 & $-$500 & \nodata & 
003904.8+402927.4 & 1.2 & 1.3 & 1.5 & 1.3 & $-$500 & \nodata \\
003906.7+403704.5 & 3.5 & 2.8 & 3.0 & 3.0 & $-$500 & \nodata & 
003909.8+402705.0 & 2.1 & 2.1 & 2.1 & 2.0 & $-$500 & \nodata \\
003910.2+403725.6 & 3.2 & 2.8 & 4.2 & 2.9 & $-$500 & \nodata & 
003914.6+404157.9 & 2.1 & 2.2 & 2.3 & 2.1 & $-$500 & \nodata \\
003916.1+403629.5 & 3.0 & 2.6 & 2.8 & 2.9 & $-$500 & \nodata & 
\textbf{003918.9+402158.4} & 0.7 & 0.6 & 0.6 & 0.5 & $-$500 & \nodata \\
003930.2+402106.4 & 3.7 & 2.6 & 3.0 & 3.5 & $-$500 & \nodata &
003933.2+402215.6 & 3.2 & 2.8 & 2.8 & 3.2 & $-$500 & \nodata \\ 
003935.2+404814.6 & 2.1 & 2.1 & 2.1 & 2.2 & $-$500 & \nodata &
003937.5+402011.5 & 3.4 & 2.9 & 2.9 & 3.2 & $-$500 & \nodata \\ 
003938.9+401921.3 & 2.5 & 2.3 & 2.4 & 2.4 & $-$500 & \nodata &
003939.1+405018.3 & 1.9 & 2.3 & 2.0 & 2.0 & $-$500 & \nodata \\ 
003939.8+402856.3 & 3.1 & 2.8 & 2.8 & 3.1 & $-$500 & \nodata &
003941.5+402133.7 & 1.7 & 2.1 & 2.2 & 2.3 & $-$500 & \nodata \\
003941.9+402045.6 & 2.0 & 2.0 & 2.1 & 2.1 & $-$500 & \nodata & 
003943.0+402039.9 & 2.1 & 2.4 & 2.5 & 2.3 & $-$500 & \nodata \\
003944.5+402030.4 & 3.3 & 2.9 & 2.5 & 2.6 & $-$500 & \nodata & 
003945.2+402058.0 & 1.9 & 2.4 & 2.4 & 2.2 & $-$500 & \nodata \\
003948.5+403113.1 & 2.0 & 1.9 & 2.0 & 2.2 & $-$500 & \nodata & 
003950.5+402305.9 & 2.2 & 2.2 & 2.2 & 2.2 & $-$500 & \nodata \\
003950.9+402252.1\tablenotemark{b} & 1.9 & 2.1 & 2.2 & 2.0 & $-$500 & \nodata & 
003951.3+405306.1 & 1.9 & 2.1 & 2.1 & 2.2 & $-$500 & \nodata \\
003954.4+403820.4\tablenotemark{a} & 3.4 & 3.5 & 3.6 & 3.6 & $-$500 & \nodata & 
003956.8+402437.6 & 1.9 & 2.2 & 2.4 & 2.1 & $-$500 & \nodata \\
004000.3+405318.6 & 2.0 & 2.2 & 2.2 & 2.1 & $-$500 & \nodata & 
004004.7+405840.9 & 3.2 & 3.2 & 3.2 & 3.3 & $-$500 & \nodata \\ 
004010.4+404517.7 & 1.9 & 2.0 & 2.3 & 2.0 & $-$500 & $-$513 & 
004020.3+403124.5 & 2.1 & 2.4 & 2.6 & 2.0 & $-$500 & \nodata \\ 
004020.5+403723.9 & 2.0 & 2.3 & 2.2 & 2.1 & $-$500 & $-$528 & 
004023.8+403904.4 & 3.4 & 3.5 & 3.7 & 3.3 & $-$500 & $-$495 \\
004026.2+403706.5 & 3.0 & 3.7 & 3.9 & 3.3 & $-$500 & \nodata & 
004030.9+404230.0 & 3.0 & 2.3 & 2.8 & 3.1 & $-$500 & $-$466 \\
004031.2+403952.0 & 2.9 & 2.4 & 2.4 & 2.5 & $-$500 & \nodata & 
004031.3+404032.8 & 2.1 & 2.0 & 2.0 & 1.6 & $-$500 & $-$453 \\
004031.7+404127.0 & 3.3 & 3.0 & 3.3 & 2.7 & $-$500 & $-$565 & 
004032.5+405127.4 & 2.0 & 2.3 & 1.9 & 1.9 & $-$500 & $-$454 \\
004032.6+403856.1 & 3.3 & 2.9 & 2.8 & 2.6 & $-$500 & $-$558 & 
004032.7+403531.2 & 2.1 & 2.4 & 2.7 & 2.0 & $-$500 & $-$538 \\
004032.7+403936.5 & 3.5 & 3.4 & 3.4 & 3.4 & $-$500 & $-$567 & 
004032.7+410045.1 & 2.0 & 2.0 & 2.0 & 2.1 & $-$500 & \nodata \\
004032.8+405540.2 & 1.9 & 2.1 & 2.2 & 2.0 & $-$500 & $-$454 & 
004032.9+403919.2 & 3.3 & 3.6 & 3.6 & 3.2 & $-$500 & $-$565 \\
004033.0+404102.8 & 3.0 & 3.7 & 3.5 & 3.1 & $-$500 & $-$576 & 
004033.3+403352.1 & 3.1 & 2.8 & 2.7 & 2.9 & $-$500 & \nodata \\
004033.8+403246.6 & 3.1 & 2.8 & 2.7 & 2.8 & $-$500 & \nodata & 
004034.7+403541.2 & 2.8 & 2.8 & 3.0 & 2.8 & $-$500 & $-$582 \\
004035.1+403701.1 & 4.0 & 3.5 & 4.0 & 3.0 & $-$500 & $-$510 & 
004035.8+403724.6 & 3.6 & 5.1 & 3.3 & 3.7 & $-$500 & \nodata \\ 
004036.0+403821.0 & 3.0 & 3.3 & 3.4 & 3.1 & $-$500 & \nodata & 
004036.1+410117.5 & 2.1 & 1.9 & 1.8 & 2.0 & $-$500 & \nodata \\ 
004036.3+403641.9 & 2.1 & 2.1 & 2.2 & 2.0 & $-$500 & $-$532 & 
004036.3+405329.3 & 1.9 & 2.0 & 2.1 & 2.1 & $-$500 & $-$575 \\
004036.8+403557.1 & 2.0 & 2.2 & 2.1 & 2.0 & $-$500 & $-$549 & 
004038.0+403514.9 & 2.9 & 3.8 & 3.3 & 3.4 & $-$500 & $-$540 \\
004038.0+404728.3 & 2.0 & 2.2 & 2.1 & 2.0 & $-$500 & \nodata & 
004038.6+403814.7 & 2.9 & 3.7 & 3.6 & 3.3 & $-$500 & \nodata \\
004038.7+403533.2 & 2.3 & 2.2 & 2.3 & 2.3 & $-$500 & $-$541 & 
004038.8+403431.0 & 3.4 & 3.5 & 3.3 & 3.0 & $-$500 & $-$504 \\
004039.4+403730.5 & 3.4 & 4.4 & 3.2 & 3.3 & $-$500 & $-$547 & 
004039.7+403457.9 & 2.2 & 2.2 & 2.2 & 2.3 & $-$500 & $-$525 \\
004040.4+402709.8\tablenotemark{b} & 2.2 & 2.2 & 2.2 & 2.0 & $-$500 & \nodata & 
004041.6+405105.0 & 2.0 & 2.0 & 1.8 & 1.7 & $-$500 & \nodata \\  
004042.1+403454.5 & 3.4 & 3.2 & 3.5 & 3.3 & $-$500 & $-$535 & 
004043.3+404321.9 & 3.1 & 3.0 & 2.9 & 2.7 & $-$500 & $-$533 \\
004043.6+403530.5 & 3.3 & 2.9 & 2.8 & 3.2 & $-$500 & $-$552 & 
004043.7+405251.5 & 2.0 & 2.0 & 2.2 & 2.1 & $-$500 & $-$445 \\
004044.2+404446.4 & 3.0 & 2.8 & 3.1 & 2.5 & $-$500 & \nodata & 
004045.7+405134.5 & 1.9 & 1.8 & 2.2 & 1.8 & $-$500 & \nodata \\
004046.4+405541.9 & 3.0 & 2.7 & 2.5 & 2.6 & $-$500 & $-$519 & 
004046.5+405606.4 & 2.8 & 2.7 & 2.5 & 2.8 & $-$500 & $-$544 \\
004047.3+405903.2 & 2.0 & 2.0 & 2.1 & 2.0 & $-$500 & $-$497 & 
004050.0+405938.5 & 1.7 & 2.3 & 2.2 & 2.0 & $-$500 & $-$502 \\
004051.6+410006.5 & 2.9 & 2.5 & 2.5 & 2.5 & $-$500 & $-$503 & 
004051.7+403602.3 & 2.1 & 2.2 & 2.3 & 2.2 & $-$500 & \nodata \\
004051.9+403249.7 & 3.4 & 3.8 & 3.4 & 2.9 & $-$500 & \nodata & 
004053.0+403218.0 & 3.8 & 3.8 & 3.5 & 3.7 & $-$500 & $-$480 \\
004055.1+403703.2 & 3.8 & 4.0 & 3.6 & 3.4 & $-$500 & $-$530 & 
004057.3+403607.0 & 2.2 & 2.3 & 2.4 & 2.2 & $-$500 & \nodata \\
004058.2+410302.3 & 1.9 & 2.2 & 2.0 & 2.1 & $-$500 & $-$476 & 
004058.3+403711.1 & 2.1 & 2.2 & 2.1 & 2.1 & $-$500 & $-$510 \\
004058.4+405325.2 & 2.2 & 2.3 & 2.4 & 2.0 & $-$500 & \nodata & 
004058.4+410217.9 & 1.9 & 2.2 & 2.1 & 1.9 & $-$500 & $-$473 \\
004058.4+410225.9 & 2.0 & 2.1 & 2.0 & 2.1 & $-$500 & $-$474 & 
004058.6+404558.0 & 2.1 & 2.0 & 1.8 & 2.1 & $-$500 & $-$506 \\
004058.6+410332.3 & 3.0 & 2.8 & 3.0 & 3.1 & $-$500 & $-$466 & 
004059.1+410233.1 & 2.1 & 2.4 & 2.3 & 2.2 & $-$500 & $-$477 \\
004059.8+403652.4 & 4.8 & 4.2 & 3.3 & 3.0 & $-$500 & $-$497 & 
004100.6+410334.0 & 3.7 & 3.0 & 3.0 & 3.2 & $-$500 & $-$461 \\
004101.6+410405.8 & 2.3 & 2.2 & 2.1 & 2.1 & $-$500 & $-$457 & 
004102.0+410254.9 & 1.7 & 1.9 & 1.9 & 1.9 & $-$500 & $-$480 \\
004102.3+410431.7 & 2.1 & 2.4 & 2.2 & 2.0 & $-$500 & $-$449 & 
004102.7+410344.5 & 2.1 & 2.1 & 2.4 & 1.9 & $-$500 & $-$456 \\
004103.1+403749.9 & 2.3 & 2.4 & 2.3 & 2.0 & $-$500 & $-$502 & 
004104.8+410534.6 & 1.8 & 1.2 & 1.2 & 1.3 & $-$500 & $-$458 \\
004107.2+410410.0 & 2.0 & 2.1 & 2.2 & 2.2 & $-$500 & $-$459 & 
004107.6+404812.5 & 3.5 & 2.9 & 2.8 & 3.0 & $-$500 & $-$526 \\
004108.6+410437.9 & 1.2 & 1.3 & 1.2 & 1.2 & $-$500 & $-$452 & 
004109.1+404852.7 & 3.0 & 3.6 & 3.9 & 3.4 & $-$500 & $-$490 \\
004109.2+404910.3 & 3.1 & 3.6 & 3.7 & 3.3 & $-$500 & $-$518 & 
004110.4+404949.5 & 3.1 & 3.0 & 3.5 & 3.0 & $-$500 & $-$555 \\
004110.6+410516.4 & 2.2 & 2.6 & 2.2 & 2.2 & $-$500 & $-$450 & 
004112.5+410609.7 & 3.4 & 2.6 & 2.9 & 2.5 & $-$300 & $-$435 \\
004113.7+403918.6 & 2.2 & 2.2 & 2.3 & 2.2 & $-$500 & $-$477 & 
004113.8+410814.6 & 2.0 & 2.1 & 2.2 & 2.0 & $-$300 & $-$422 \\
004113.9+410736.1 & 2.2 & 2.0 & 2.2 & 2.0 & $-$300 & $-$431 & 
004114.8+410923.7 & 3.8 & 3.0 & 3.3 & 3.4 & $-$300 & $-$413 \\
004115.9+404011.6 & 3.1 & 2.8 & 2.6 & 2.7 & $-$500 & $-$448 & 
004119.1+404857.4 & 3.6 & 3.1 & 3.2 & 3.4 & $-$500 & $-$473 \\
004119.5+411948.8 & 3.9 & 3.3 & 3.1 & 3.0 & $-$300 & \nodata &  
004120.0+410821.5 & 3.0 & 2.7 & 2.8 & 2.6 & $-$300 & \nodata \\ 
004120.9+403414.0 & 2.4 & 2.7 & 2.6 & 2.5 & $-$500 & \nodata & 
004121.2+411947.8 & 2.5 & 2.7 & 2.5 & 2.5 & $-$300 & \nodata \\ 
\textbf{004121.7+404947.7} & 1.4 & 1.5 & 1.6 & 1.4 & $-$500 & $-$520 & 
004123.2+405000.6 & 3.4 & 3.3 & 2.9 & 3.0 & $-$500 & $-$518 \\
004124.1+411124.1 & 3.1 & 3.7 & 3.7 & 3.7 & $-$300 & $-$398 & 
004124.8+411154.6 & 3.4 & 2.9 & 2.7 & 2.6 & $-$300 & $-$407 \\
004125.4+404200.4 & 3.7 & 3.2 & 3.1 & 3.3 & $-$500 & $-$448 & 
004126.1+404959.1 & 3.7 & 3.2 & 3.3 & 3.5 & $-$500 & $-$502 \\
004126.5+411206.9 & 3.0 & 3.0 & 2.8 & 2.8 & $-$300 & $-$400 & 
004127.3+404242.7 & 2.0 & 2.3 & 2.3 & 2.2 & $-$500 & $-$443 \\
004128.1+404155.2 & 2.0 & 2.4 & 2.1 & 2.2 & $-$500 & $-$462 & 
004128.1+411222.6 & 3.6 & 4.0 & 4.4 & 3.2 & $-$300 & $-$389 \\
004129.2+411242.8 & 4.2 & 4.2 & 4.1 & 3.6 & $-$300 & $-$417 & 
004129.3+404218.9 & 2.2 & 2.3 & 2.2 & 2.2 & $-$500 & $-$440 \\
004129.5+411006.3 & 4.0 & 3.9 & 4.4 & 3.8 & $-$300 & \nodata & 
004129.8+405059.5 & 2.2 & 1.6 & 1.7 & 1.7 & $-$500 & $-$463 \\
004129.8+412211.1\tablenotemark{a} & 3.1 & 3.0 & 2.8 & 2.9 & $-$300 & \nodata & 
004130.3+410501.7 & 3.1 & 3.0 & 3.4 & 3.1 & $-$500 & $-$477 \\
004131.9+411331.5 & 3.8 & 4.4 & 4.1 & 4.1 & $-$300 & $-$401 & 
004135.7+405009.3 & 2.4 & 2.8 & 2.8 & 2.4 & $-$500 & \nodata \\
004136.9+403805.6 & 2.2 & 2.4 & 2.2 & 2.3 & $-$500 & \nodata & 
004137.0+405142.5 & 3.3 & 3.3 & 3.0 & 3.5 & $-$500 & $-$432 \\
004138.6+404401.2 & 3.4 & 3.1 & 3.0 & 3.1 & $-$500 & $-$426 & 
004141.3+411916.7 & 2.4 & 2.3 & 2.6 & 2.5 & $-$300 & \nodata \\
004143.6+410840.1 & 2.0 & 2.0 & 2.1 & 2.0 & $-$500 & $-$404 & 
004144.6+411658.1 & 3.9 & 4.4 & 4.1 & 3.9 & $-$300 & $-$365 \\
004145.0+404746.4 & 2.2 & 2.5 & 2.4 & 2.4 & $-$500 & \nodata & 
004146.7+411846.6 & 3.3 & 2.8 & 3.1 & 3.0 & $-$300 & $-$308 \\
004147.4+411942.4 & 3.9 & 4.3 & 4.4 & 4.1 & $-$300 & $-$359 & 
004148.2+411903.8 & 3.6 & 3.2 & 4.5 & 3.1 & $-$300 & $-$358 \\
004149.6+411953.8 & 2.4 & 2.7 & 2.5 & 2.3 & $-$300 & $-$387 & 
004151.3+412500.7 & 2.8 & 2.6 & 2.7 & 2.4 & $-$300 & \nodata \\
004151.6+404620.5 & 2.0 & 2.4 & 2.1 & 2.1 & $-$500 & \nodata &  
004151.9+412442.1 & 3.3 & 3.1 & 3.3 & 3.6 & $-$300 & \nodata \\
004154.5+404718.9 & 2.2 & 2.3 & 2.2 & 2.1 & $-$500 & $-$414 & 
004159.4+405720.8 & 2.1 & 1.7 & 1.7 & 1.8 & $-$500 & \nodata \\
004200.6+404747.8 & 2.2 & 2.5 & 2.4 & 2.3 & $-$300 & \nodata & 
004200.9+405217.2 & 2.1 & 2.2 & 2.2 & 2.1 & $-$500 & $-$407 \\
004201.5+404115.7\tablenotemark{a} & 2.3 & 2.4 & 2.5 & 2.3 & $-$300 & \nodata & 
004202.4+412436.0 & 1.1 & 1.1 & 1.2 & 1.3 & $-$300 & $-$323 \\
004202.9+412232.4 & 1.1 & 1.3 & 1.2 & 1.1 & $-$300 & $-$353 & 
004203.9+404907.1 & 2.6 & 2.1 & 2.1 & 2.4 & $-$300 & \nodata \\ 
004204.9+404936.6 & 2.2 & 2.4 & 2.2 & 2.5 & $-$300 & \nodata & 
004206.7+405621.5 & 2.0 & 2.3 & 2.0 & 2.0 & $-$500 & \nodata \\
004208.5+405720.2 & 2.0 & 2.0 & 2.1 & 2.1 & $-$500 & \nodata & 
004208.5+412409.8 & 2.2 & 2.2 & 2.1 & 2.1 & $-$300 & $-$330 \\
004208.7+405052.1 & 2.2 & 2.1 & 2.3 & 2.0 & $-$300 & $-$444 & 
004208.8+412639.9 & 3.2 & 2.8 & 2.9 & 2.7 & $-$300 & \nodata \\
004208.9+412329.8\tablenotemark{a} & 2.3 & 2.7 & 2.2 & 2.1 & $-$300 & $-$335 & 
004209.0+412442.3 & 2.3 & 2.3 & 2.3 & 2.3 & $-$300 & $-$291 \\
004209.5+412705.5 & 2.1 & 2.4 & 2.3 & 2.0 & $-$300 & \nodata & 
004209.5+412832.3 & 3.3 & 2.9 & 3.1 & 2.6 & $-$300 & \nodata \\
004209.8+412412.2 & 2.2 & 2.3 & 2.2 & 2.3 & $-$300 & $-$331 & 
004210.3+412529.3 & 2.1 & 2.7 & 2.4 & 2.5 & $-$300 & $-$345 \\
004210.7+412322.3\tablenotemark{a} & 2.2 & 2.0 & 2.3 & 2.4 & $-$300 & $-$302 & 
004211.2+412442.6 & 2.4 & 2.1 & 2.3 & 2.3 & $-$300 & $-$295 \\
004211.6+411909.4 & 2.7 & 2.8 & 2.4 & 2.6 & $-$300 & \nodata & 
004212.3+412415.7 & 2.9 & 2.7 & 2.7 & 2.6 & $-$300 & $-$330 \\
004213.8+405117.7 & 3.9 & 3.4 & 3.7 & 3.5 & $-$300 & $-$406 & 
004214.8+412508.9 & 2.8 & 3.0 & 2.8 & 2.6 & $-$300 & $-$316 \\
004218.1+412631.1 & 2.4 & 2.1 & 2.6 & 2.5 & $-$300 & $-$274 & 
004218.7+412751.8 & 2.7 & 2.7 & 2.5 & 2.6 & $-$300 & $-$330 \\
004220.6+412749.0 & 3.1 & 2.6 & 2.6 & 2.7 & $-$300 & $-$321 & 
004221.7+412827.6 & 2.8 & 4.0 & 3.2 & 2.7 & $-$300 & $-$263 \\
004224.8+412758.7 & 2.9 & 3.0 & 3.0 & 2.8 & $-$300 & $-$307 & 
004225.9+412831.9 & 2.9 & 2.7 & 2.7 & 2.9 & $-$300 & $-$302 \\
004226.1+410548.2\tablenotemark{a} & 3.8 & 3.0 & 2.9 & 2.9 & $-$500 & $-$494 & 
004226.4+412811.2 & 3.4 & 2.5 & 2.3 & 3.1 & $-$300 & $-$290 \\
004227.6+412019.6 & 2.1 & 2.3 & 2.1 & 2.0 & $-$300 & $-$226 & 
004227.9+413258.5\tablenotemark{a} & 2.5 & 2.7 & 2.7 & 2.5 & $-$300 & \nodata \\
004228.1+405657.7\tablenotemark{a} & 2.7 & 2.3 & 2.4 & 2.3 & $-$300 & $-$427 & 
004228.3+412911.4\tablenotemark{a} & 3.1 & 2.9 & 2.6 & 3.1 & $-$300 & $-$297 \\
004228.4+412852.4\tablenotemark{a} & 1.2 & 1.3 & 1.5 & 1.4 & $-$300 & $-$285 & 
004229.8+410550.6 & 2.9 & 2.6 & 2.5 & 2.6 & $-$300 & \nodata \\
004230.1+412904.0\tablenotemark{a} & 2.8 & 2.5 & 2.6 & 2.6 & $-$300 & $-$282 & 
004230.3+412935.9 & 4.0 & 3.0 & 3.2 & 3.3 & $-$300 & $-$277 \\
004230.9+405714.6\tablenotemark{a} & 3.0 & 2.8 & 2.9 & 3.2 & $-$300 & $-$396 & 
004232.1+412936.5 & 2.6 & 2.5 & 2.5 & 2.5 & $-$300 & $-$264 \\
004232.3+413008.7 & 2.7 & 2.4 & 2.7 & 2.6 & $-$300 & $-$277 & 
004232.7+411143.6 & 2.0 & 2.1 & 2.1 & 2.1 & $-$300 & \nodata \\
004234.2+413007.3 & 3.0 & 2.4 & 3.0 & 2.6 & $-$300 & $-$267 & 
004235.0+404838.1 & 2.2 & 2.3 & 2.4 & 2.2 & $-$300 & \nodata \\
004235.3+413224.4 & 2.4 & 2.6 & 2.5 & 2.7 & $-$300 & $-$265 & 
004235.6+413149.0 & 3.2 & 2.8 & 2.6 & 2.8 & $-$300 & $-$316 \\
004236.4+413308.7 & 2.6 & 2.6 & 2.5 & 2.7 & $-$300 & $-$245 & 
004236.9+410158.0 & 2.5 & 2.4 & 2.4 & 2.4 & $-$300 & $-$419 \\
004237.4+414158.3\tablenotemark{a} & 1.8 & 2.1 & 2.1 & 2.3 & $-$300 & \nodata & 
004238.6+413150.5 & 3.7 & 2.9 & 2.9 & 2.8 & $-$300 & $-$268 \\
004238.9+413135.6 & 2.9 & 2.9 & 2.6 & 3.3 & $-$300 & $-$276 & 
004240.1+410222.7 & 3.5 & 2.8 & 2.7 & 2.6 & $-$300 & $-$458 \\
004240.9+405910.8 & 2.4 & 2.6 & 2.4 & 2.3 & $-$300 & $-$367 & 
004241.3+412246.7 & 4.1 & 3.1 & 3.2 & 3.4 & $-$300 & \nodata \\  
004241.7+411435.0\tablenotemark{a} & 3.4 & 4.0 & 3.8 & 3.5 & $-$300 & \nodata & 
004241.7+413245.4 & 2.5 & 2.7 & 2.5 & 2.5 & $-$300 & $-$265 \\
004241.9+405155.2\tablenotemark{a} & 3.6 & 3.3 & 3.5 & 3.7 & $-$300 & \nodata & 
004242.1+410303.0 & 2.7 & 2.4 & 2.7 & 2.5 & $-$300 & $-$348 \\
004242.5+410001.4 & 3.0 & 3.0 & 2.9 & 3.2 & $-$300 & $-$372 & 
004242.5+413155.2 & 3.9 & 3.2 & 3.4 & 3.3 & $-$300 & $-$266 \\
004242.6+411722.5\tablenotemark{a} & 4.0 & 3.7 & 3.6 & 3.3 & $-$300 & \nodata & 
004242.9+413159.8 & 3.7 & 3.6 & 3.6 & 3.8 & $-$300 & $-$269 \\
004244.1+413259.2 & 3.4 & 2.5 & 2.9 & 2.9 & $-$300 & $-$258 & 
004244.4+411608.5\tablenotemark{a} & 3.7 & 3.4 & 3.2 & 3.6 & $-$300 & \nodata \\
004244.9+413338.6 & 2.3 & 2.7 & 2.7 & 2.5 & $-$300 & $-$245 & 
004245.0+405448.3 & 2.2 & 2.3 & 2.6 & 2.5 & $-$300 & \nodata \\
004245.2+413316.7 & 2.6 & 2.6 & 2.6 & 2.6 & $-$300 & $-$257 & 
004245.3+411656.9\tablenotemark{a} & 3.6 & 3.6 & 3.8 & 4.0 & $-$300 & \nodata \\
004246.2+410111.4 & 2.4 & 2.3 & 2.4 & 2.1 & $-$300 & $-$428 & 
004246.8+414447.0 & 2.2 & 2.1 & 2.0 & 2.0 & $-$300 & \nodata \\
004247.0+411618.4\tablenotemark{a} & 3.8 & 3.9 & 3.7 & 3.9 & $-$300 & \nodata & 
004247.0+413333.0 & 4.0 & 3.1 & 3.3 & 3.1 & $-$300 & $-$248 \\
004247.5+413131.1 & 2.6 & 2.4 & 2.5 & 2.5 & $-$300 & \nodata & 
004247.9+413400.5 & 3.1 & 3.0 & 3.1 & 2.9 & $-$300 & $-$254 \\
004248.2+411651.7\tablenotemark{a} & 3.6 & 3.9 & 4.0 & 3.6 & $-$300 & \nodata & 
004249.1+411554.6\tablenotemark{a} & 3.8 & 3.6 & 4.0 & 3.5 & $-$300 & \nodata \\
004249.1+411945.9\tablenotemark{a} & 3.1 & 2.8 & 2.4 & 2.6 & $-$300 & \nodata & 
004249.1+413440.0 & 2.3 & 2.7 & 2.6 & 2.5 & $-$300 & $-$244 \\
004249.3+412507.5 & 3.2 & 2.6 & 2.7 & 2.7 & $-$300 & $-$171 & 
004251.0+413507.8 & 3.2 & 2.6 & 3.0 & 3.0 & $-$300 & $-$242 \\
004252.3+410014.8 & 2.1 & 2.0 & 2.0 & 1.9 & $-$300 & \nodata & 
004252.4+410120.7 & 2.8 & 2.9 & 2.9 & 2.4 & $-$300 & $-$331 \\
004253.0+413526.7 & 2.6 & 2.4 & 2.4 & 2.4 & $-$300 & $-$251 & 
004253.5+413516.2 & 2.4 & 2.4 & 2.3 & 2.3 & $-$300 & $-$254 \\
004254.4+405832.8 & 2.3 & 2.4 & 2.6 & 3.0 & $-$300 & \nodata & 
004256.9+413728.1 & 1.6 & 1.7 & 1.8 & 1.4 & $-$300 & $-$304 \\
004258.2+410015.9 & 3.1 & 2.5 & 2.5 & 2.8 & $-$300 & \nodata & 
004258.8+413456.2 & 1.9 & 2.0 & 2.0 & 2.0 & $-$300 & $-$212 \\
004259.1+413741.3 & 1.9 & 2.2 & 2.1 & 1.9 & $-$300 & $-$268 & 
004259.4+413722.5 & 2.0 & 2.2 & 2.2 & 1.8 & $-$300 & $-$273 \\
004259.4+413732.1 & 2.0 & 2.1 & 2.1 & 1.8 & $-$300 & $-$265 & 
004300.0+413526.2 & 2.0 & 2.0 & 2.1 & 1.9 & $-$300 & $-$184 \\
004300.0+413654.2 & 2.5 & 2.4 & 2.3 & 2.1 & $-$300 & $-$248 & 
004301.0+413627.9 & 1.4 & 1.3 & 1.3 & 1.2 & $-$300 & $-$248 \\
004301.5+413717.2 & 2.8 & 2.6 & 2.7 & 2.6 & $-$300 & $-$244 & 
004301.9+413655.2 & 1.8 & 2.1 & 2.2 & 1.9 & $-$300 & $-$234 \\
004302.5+413740.5 & 3.5 & 3.3 & 3.2 & 3.7 & $-$300 & $-$241 & 
004302.5+414910.5 & 2.1 & 2.0 & 2.2 & 2.0 & $-$300 & \nodata \\
004303.4+413719.3 & 1.8 & 2.0 & 2.0 & 1.8 & $-$300 & $-$245 & 
004304.3+413739.5 & 2.0 & 2.3 & 2.2 & 1.9 & $-$300 & $-$232 \\
004304.8+410554.0 & 2.7 & 2.3 & 2.6 & 2.5 & $-$300 & $-$374 & 
004305.7+413749.8 & 2.0 & 1.9 & 1.8 & 1.9 & $-$300 & $-$220 \\
004306.7+410213.4 & 2.2 & 2.4 & 2.4 & 2.4 & $-$300 & \nodata & 
004306.9+413807.1 & 3.3 & 2.6 & 2.8 & 2.9 & $-$300 & $-$170 \\
004308.2+410156.8 & 1.3 & 1.5 & 1.5 & 1.5 & $-$300 & \nodata & 
004309.7+413849.3 & 1.8 & 1.9 & 2.3 & 2.1 & $-$300 & $-$216 \\
004310.0+413751.6\tablenotemark{b} & 1.8 & 2.0 & 2.0 & 2.0 & $-$300 & $-$226 & 
004310.5+410426.8 & 2.3 & 2.4 & 2.6 & 2.6 & $-$300 & \nodata \\ 
004311.1+413743.3 & 1.8 & 2.2 & 2.1 & 2.2 & $-$300 & $-$174 & 
004311.3+410459.5 & 3.6 & 3.3 & 3.5 & 3.7 & $-$300 & $-$380 \\
004311.6+411245.5 & 2.5 & 2.4 & 2.3 & 2.9 & $-$300 & \nodata & 
004312.4+410125.2 & 2.3 & 2.6 & 2.5 & 2.3 & $-$300 & \nodata \\
004312.5+413747.4 & 1.8 & 2.1 & 2.0 & 2.0 & $-$300 & \nodata &  
004312.7+410531.5 & 1.2 & 1.3 & 1.3 & 1.4 & $-$300 & $-$319 \\
004313.2+410632.4 & 2.3 & 2.5 & 2.3 & 2.6 & $-$300 & $-$339 & 
004314.0+413906.3 & 1.9 & 2.5 & 2.4 & 1.9 & $-$300 &  \nodata \\ 
004314.2+410033.9\tablenotemark{a} & 2.4 & 2.3 & 2.3 & 2.3 & $-$300 & \nodata & 
004315.2+414947.4 & 2.0 & 2.6 & 2.4 & 2.3 & $-$300 & \nodata \\
004317.9+410252.8 & 1.5 & 1.4 & 1.4 & 1.5 & $-$300 & \nodata &  
004320.1+410611.1 & 1.4 & 1.2 & 1.2 & 1.3 & $-$300 & $-$328 \\
004320.8+414038.5 & 2.1 & 2.3 & 2.2 & 2.1 & $-$300 & $-$224 & 
004321.7+414033.2 & 2.3 & 2.3 & 2.2 & 2.5 & $-$300 & $-$261 \\
004322.0+414116.5 & 2.1 & 2.2 & 2.1 & 2.0 & $-$300 & $-$193 & 
004324.1+414124.7 & 2.0 & 1.6 & 1.8 & 1.8 & $-$300 & \nodata \\ 
004324.3+414418.7 & 1.9 & 1.7 & 1.7 & 1.9 & $-$300 & \nodata & 
004324.4+410802.9 & 2.5 & 2.4 & 2.3 & 2.1 & $-$300 & $-$312 \\
004325.6+410206.4\tablenotemark{a} & 4.4 & 3.8 & 3.2 & 3.9 & $-$300 & \nodata & 
004326.4+410508.4 & 2.4 & 2.3 & 2.3 & 2.1 & $-$300 & \nodata \\
004328.2+414122.1 & 1.6 & 1.8 & 1.9 & 2.1 & $-$300 & \nodata & 
004328.6+411818.0 & 3.9 & 4.0 & 3.9 & 3.9 & $-$300 & $-$299 \\
004329.2+414848.0\tablenotemark{b} & 1.2 & 1.4 & 1.5 & 1.4 & $-$300 & $-$226 & 
004330.4+412757.0 & 2.6 & 2.6 & 2.5 & 2.3 & $-$100 & \nodata \\
004330.4+414432.5 & 2.4 & 2.2 & 2.1 & 2.4 & $-$300 & $-$231 & 
004331.2+414222.9 & 2.1 & 2.2 & 2.2 & 2.0 & $-$300 & $-$215 \\
004331.3+414243.6 & 1.8 & 2.0 & 1.9 & 2.2 & $-$300 & $-$248 & 
004332.1+414251.8 & 1.2 & 1.2 & 1.3 & 1.5 & $-$300 & $-$242 \\
004332.4+414227.5 & 1.7 & 1.7 & 1.9 & 2.0 & $-$300 & $-$211 & 
004332.5+410907.0\tablenotemark{a} & 2.6 & 3.1 & 2.7 & 2.7 & $-$300 & $-$320 \\
004333.6+411432.3 & 2.8 & 2.7 & 2.4 & 2.7 & $-$300 & $-$306 & 
004334.9+410953.6 & 3.8 & 3.2 & 3.4 & 3.9 & $-$300 & $-$308 \\
004338.7+411222.1 & 2.4 & 2.6 & 2.6 & 2.7 & $-$300 & $-$243 & 
004339.1+411018.4 & 2.5 & 2.5 & 2.8 & 3.1 & $-$300 & $-$315 \\
004339.3+411001.1 & 2.8 & 2.9 & 2.7 & 2.6 & $-$300 & $-$306 & 
004339.4+412229.2\tablenotemark{a} & 2.3 & 2.3 & 2.3 & 2.3 & $-$300 &  \nodata \\ 
004339.7+414534.9 & 2.0 & 2.3 & 2.1 & 2.3 & $-$300 & \nodata &  
004340.8+411152.7 & 2.6 & 2.4 & 2.7 & 2.6 & $-$300 & $-$280 \\
004341.5+414224.3\tablenotemark{a} & 2.5 & 2.5 & 2.6 & 2.2 & $-$300 & \nodata & 
004341.6+411135.3 & 2.8 & 2.5 & 2.5 & 2.5 & $-$300 & $-$307 \\
004341.7+412302.9 & 2.5 & 2.7 & 2.6 & 2.3 & $-$300 & $-$173 & 
004341.7+414519.4 & 2.0 & 1.6 & 1.6 & 1.7 & $-$300 & $-$156 \\
004341.7+415313.0\tablenotemark{a} & 2.0 & 2.4 & 2.2 & 2.0 & $-$300 & \nodata & 
004343.4+414521.6 & 2.1 & 2.1 & 2.3 & 2.2 & $-$300 & $-$142 \\
\textbf{004343.9+411137.6} & 1.3 & 1.2 & 1.4 & 1.1 & $-$300 & $-$295 & 
004344.6+412321.3 & 2.4 & 2.4 & 2.4 & 2.6 & $-$300 & $-$168 \\
004346.3+414418.5 & 3.8 & 3.3 & 3.3 & 3.5 & $-$300 & $-$194 & 
004346.8+411239.7 & 2.9 & 2.6 & 2.4 & 2.7 & $-$300 & $-$262 \\
004348.1+411133.2 & 3.2 & 2.9 & 3.0 & 3.0 & $-$300 & $-$297 & 
004349.0+415657.7 & 3.6 & 4.1 & 4.1 & 4.1 & $-$300 & $-$246 \\
004349.4+411053.8 & 2.4 & 2.2 & 2.1 & 2.1 & $-$300 & \nodata & 
004351.4+414706.2 & 3.5 & 3.7 & 3.8 & 4.0 & $-$300 & \nodata \\ 
004351.4+415718.7 & 2.1 & 2.4 & 2.4 & 1.9 & $-$300 & \nodata &  
004352.5+412524.1 & 2.8 & 2.6 & 2.7 & 3.1 & $-$100 & $-$184 \\
004352.5+414858.8 & 2.3 & 2.4 & 2.5 & 2.1 & $-$300 & \nodata & 
004353.9+415743.8 & 1.8 & 2.2 & 2.2 & 2.0 & $-$300 & \nodata \\ 
004354.8+414715.6 & 2.9 & 2.5 & 2.6 & 2.5 & $-$300 & \nodata & 
004354.9+412603.6 & 3.2 & 2.4 & 2.7 & 2.7 & $-$100 & $-$154 \\
004355.1+411433.1 & 2.4 & 2.2 & 2.4 & 2.4 & $-$300 & \nodata & 
004355.2+412650.8 & 3.0 & 3.2 & 2.8 & 2.8 & $-$100 & $-$169 \\
004355.8+411211.6\tablenotemark{b} & 4.1 & 3.5 & 3.0 & 3.9 & $-$300 & $-$249 & 
004356.6+412629.6 & 2.4 & 2.1 & 2.0 & 1.8 & $-$100 & $-$145 \\
004356.8+414831.6 & 3.6 & 3.4 & 3.5 & 3.8 & $-$300 & $-$169 & 
004357.7+414854.0 & 3.1 & 2.6 & 2.8 & 2.6 & $-$300 & $-$154 \\
004358.2+414726.9 & 3.0 & 2.3 & 2.6 & 2.3 & $-$300 & \nodata &  
004358.7+414837.5 & 3.0 & 2.5 & 2.6 & 2.4 & $-$300 & $-$116 \\
004358.9+411742.1 & 2.6 & 2.4 & 2.5 & 2.1 & $-$300 & $-$291 & 
004401.5+414909.6 & 3.8 & 3.6 & 3.6 & 3.3 & $-$300 & $-$166 \\
004403.0+414954.7 & 3.1 & 2.9 & 2.8 & 2.4 & $-$300 & $-$168 &
004403.4+411708.2 & 2.2 & 2.3 & 2.3 & 2.5 & $-$300 & $-$291 \\ 
004403.9+413414.8\tablenotemark{a} & 3.4 & 2.7 & 2.8 & 2.6 & $-$100 & \nodata & 
004404.2+412107.8 & 2.5 & 2.3 & 2.2 & 2.6 & $-$300 & $-$266 \\
004404.9+415016.1 & 2.8 & 2.7 & 2.6 & 2.9 & $-$300 & $-$181 & 
004405.2+412718.2 & 2.8 & 2.4 & 2.8 & 2.9 & $-$100 & $-$128 \\
004405.7+411719.7 & 2.3 & 2.9 & 2.5 & 2.4 & $-$300 & $-$255 & 
004406.4+412745.0 & 2.9 & 2.6 & 2.8 & 2.4 & $-$100 & $-$141 \\
004407.0+412759.3 & 3.3 & 2.8 & 3.0 & 3.0 & $-$100 & $-$144 & 
004407.8+412110.7 & 3.0 & 2.7 & 2.7 & 2.6 & $-$300 & $-$248 \\
004409.2+413331.9 & 2.4 & 2.3 & 2.3 & 2.3 & $-$100 & $-$92 & 
\textbf{004409.5+411856.6} & 1.2 & 1.3 & 1.3 & 1.3 & $-$300 & $-$254 \\
004410.5+420247.5\tablenotemark{a} & 2.9 & 2.7 & 3.0 & 2.6 & $-$100 & \nodata & 
004410.6+411653.4 & 2.2 & 2.4 & 2.4 & 2.3 & $-$300 & \nodata \\ 
004411.0+413206.3 & 2.1 & 2.4 & 2.5 & 2.5 & $-$100 & $-$127 & 
004411.8+414747.5 & 2.2 & 2.3 & 2.1 & 1.8 & $-$100 & \nodata \\
004411.9+413356.4 & 2.1 & 2.3 & 2.2 & 2.3 & $-$100 & $-$80 & 
004412.1+413320.5 & 2.8 & 3.1 & 2.7 & 2.8 & $-$100 & $-$103 \\
004413.7+413413.5 & 2.2 & 2.2 & 2.4 & 2.2 & $-$100 & $-$80 & 
004414.4+411742.3 & 2.4 & 2.7 & 2.4 & 2.3 & $-$300 & $-$266 \\
004414.6+412840.3 & 2.3 & 2.5 & 2.5 & 2.7 & $-$100 & $-$113 & 
004415.3+411905.7 & 2.9 & 2.5 & 3.8 & 2.7 & $-$300 & $-$236 \\
004415.9+411717.6 & 2.1 & 2.2 & 2.4 & 2.2 & $-$300 & $-$252 & 
004416.0+414950.7 & 3.1 & 2.7 & 3.0 & 2.7 & $-$100 & \nodata \\
004416.1+412105.4 & 2.3 & 2.4 & 2.5 & 2.2 & $-$300 & $-$244 & 
004416.3+411730.9 & 2.4 & 2.6 & 2.4 & 2.4 & $-$300 & $-$182 \\
004416.7+412444.1 & 2.1 & 2.3 & 2.2 & 2.1 & $-$300 & $-$277 & 
004418.2+413406.6 & 2.8 & 2.8 & 2.8 & 2.8 & $-$100 & $-$152 \\
004419.2+411930.9 & 2.9 & 2.5 & 2.5 & 2.7 & $-$300 & $-$242 & 
004419.3+412247.0 & 2.4 & 2.6 & 2.5 & 2.5 & $-$300 & $-$275 \\
004419.9+412201.2 & 3.2 & 2.7 & 2.5 & 2.7 & $-$300 & $-$272 & 
004420.2+415101.5 & 3.4 & 2.5 & 2.6 & 2.8 & $-$100 & \nodata \\ 
004420.7+411751.0 & 2.3 & 2.6 & 2.7 & 2.2 & $-$300 & \nodata &  
004420.9+411835.7 & 2.5 & 2.5 & 2.5 & 2.4 & $-$300 & $-$254 \\
004422.8+412529.1 & 2.1 & 2.3 & 2.2 & 2.1 & $-$300 & \nodata & 
004423.0+412050.9 & 3.3 & 2.9 & 3.0 & 2.7 & $-$300 & $-$245 \\
004423.3+413842.6 & 2.8 & 2.6 & 3.0 & 2.7 & $-$100 & \nodata & 
004423.7+412437.3 & 3.3 & 2.5 & 2.5 & 2.6 & $-$300 & $-$233 \\
004424.1+412117.3 & 2.9 & 2.8 & 2.7 & 2.9 & $-$300 & $-$218 & 
004424.2+414918.9 & 3.2 & 2.4 & 2.5 & 2.7 & $-$100 & $-$102 \\
004424.4+415120.5 & 1.7 & 2.0 & 2.1 & 2.0 & $-$100 & $-$164 & 
004424.9+413739.1 & 2.1 & 2.1 & 1.9 & 1.8 & $-$100 & $-$66 \\
004425.0+414942.6 & 1.8 & 2.2 & 2.1 & 1.9 & $-$100 & \nodata &  
004425.4+415006.1 & 1.8 & 2.2 & 2.1 & 2.1 & $-$100 & \nodata \\ 
004426.2+412054.1 & 2.4 & 2.1 & 2.1 & 2.0 & $-$300 & $-$226 & 
004426.7+412729.3 & 2.4 & 2.3 & 2.2 & 2.2 & $-$300 & $-$217 \\
004427.5+413529.8 & 3.0 & 2.5 & 2.7 & 2.5 & $-$100 & \nodata & 
004429.1+412334.0 & 3.2 & 2.5 & 2.8 & 2.6 & $-$300 & $-$228 \\
004429.6+412138.9 & 2.1 & 2.1 & 2.1 & 2.2 & $-$300 & $-$268 & 
004429.6+415133.5 & 1.9 & 2.0 & 1.9 & 2.0 & $-$100 & $-$113 \\
004430.2+415242.7 & 2.0 & 2.1 & 2.3 & 2.0 & $-$100 & $-$107 & 
\textbf{004430.5+415154.8} & 1.1 & 1.1 & 1.1 & 1.2 & $-$100 & $-$96 \\
004431.1+415110.2 & 3.0 & 2.6 & 2.9 & 3.0 & $-$100 & $-$104 & 
004431.1+415638.2 & 1.9 & 2.0 & 2.1 & 1.9 & $-$100 & $-$94 \\
004431.9+412233.3 & 2.9 & 2.6 & 2.7 & 2.6 & $-$300 & $-$196 & 
004431.9+412400.1 & 2.2 & 2.2 & 2.0 & 4.5 & $-$300 & $-$219 \\ 
004432.6+412518.7 & 2.2 & 2.3 & 2.3 & 2.4 & $-$300 & $-$172 & 
004433.8+415249.7\tablenotemark{b} & 2.3 & 2.1 & 1.9 & 2.1 & $-$100 & $-$90 \\
004435.6+415606.9\tablenotemark{b} & 3.1 & 2.6 & 2.6 & 2.8 & $-$100 & \nodata &  
004436.7+412445.1 & 2.3 & 2.9 & 2.6 & 2.5 & $-$300 & $-$189 \\ 
004437.3+415350.2 & 1.9 & 2.1 & 2.3 & 1.9 & $-$100 & \nodata &   
004437.7+415259.8 & 2.1 & 2.4 & 2.3 & 2.4 & $-$100 & $-$101 \\
004437.9+415154.0 & 3.8 & 2.5 & 2.7 & 2.6 & $-$100 & $-$137 & 
004438.5+412511.1 & 3.7 & 3.4 & 3.3 & 3.6 & $-$300 & $-$211 \\
004439.4+415251.3 & 2.2 & 2.4 & 2.2 & 2.3 & $-$100 & $-$107 & 
004440.3+414923.9 & 1.9 & 2.2 & 2.2 & 2.0 & $-$100 & \nodata \\
004441.5+415312.7 & 1.9 & 2.0 & 2.2 & 1.9 & $-$100 & $-$103 & 
004441.7+412659.6 & 2.1 & 2.4 & 2.1 & 2.2 & $-$300 & $-$208 \\
004442.7+415341.1 & 1.3 & 1.4 & 1.4 & 1.3 & $-$100 & $-$98 & 
004443.9+412758.0 & 3.9 & 3.3 & 3.9 & 3.6 & $-$300 & $-$186 \\
004444.1+415359.0 & 3.3 & 2.5 & 2.6 & 2.7 & $-$100 & $-$59 & 
004444.8+412839.9 & 2.6 & 2.4 & 2.6 & 2.6 & $-$300 & $-$165 \\
004447.1+415657.7 & 2.5 & 2.5 & 2.3 & 2.2 & $-$100 & \nodata & 
004447.6+412641.5 & 2.5 & 2.5 & 2.5 & 2.5 & $-$300 & \nodata \\  
004448.1+415307.3 & 2.3 & 2.5 & 2.4 & 2.4 & $-$100 & $-$91 & 
004448.4+412254.2 & 3.1 & 3.0 & 2.9 & 2.7 & $-$300 & \nodata \\
004448.6+415343.6 & 2.3 & 2.5 & 2.5 & 2.3 & $-$100 & $-$87 & 
004450.6+415608.3 & 2.3 & 2.3 & 2.3 & 2.0 & $-$100 & $-$42 \\
004450.9+412909.2 & 2.2 & 2.4 & 2.3 & 2.3 & $-$300 & $-$179 & 
004451.8+415423.7 & 1.2 & 1.3 & 1.4 & 1.3 & $-$100 & $-$131 \\ 
004452.7+415309.1 & 1.3 & 1.5 & 1.4 & 1.5 & $-$100 & $-$62 & 
004452.8+415457.5 & 2.3 & 2.1 & 2.0 & 2.3 & $-$100 & $-$95 \\
004453.0+415340.3 & 3.0 & 2.9 & 2.8 & 3.0 & $-$100 & $-$101 & 
004454.4+420327.4 & 1.8 & 2.0 & 2.2 & 2.1 & $-$100 & \nodata \\
004456.1+412918.2 & 3.2 & 2.9 & 2.8 & 2.7 & $-$300 & $-$174 & 
004456.2+413124.1 & 2.5 & 2.3 & 2.0 & 2.1 & $-$300 & $-$153 \\
004457.2+415524.0 & 3.1 & 2.7 & 2.6 & 2.6 & $-$100 & $-$72 & 
004457.3+413141.8 & 2.1 & 2.4 & 2.4 & 1.9 & $-$300 & \nodata \\  
004458.0+414034.7 & 1.3 & 1.6 & 1.4 & 1.3 & $-$100 & \nodata &   
004458.1+420008.6 & 2.3 & 2.2 & 2.3 & 2.3 & $-$100 & $-$114 \\
004458.3+415906.9 & 2.4 & 2.4 & 2.4 & 2.2 & $-$100 & \nodata & 
004458.7+415536.1 & 2.1 & 1.9 & 1.7 & 3.2 & $-$100 & $-$81 \\
004459.1+413233.8 & 4.2 & 3.3 & 4.8 & 3.3 & $-$300 & $-$166 & 
004459.1+414058.5 & 3.1 & 2.7 & 2.8 & 2.9 & $-$100 & \nodata \\  
004459.3+413139.2 & 2.9 & 2.2 & 2.4 & 2.5 & $-$300 & $-$180 & 
004459.5+415510.4 & 2.4 & 2.6 & 2.3 & 2.4 & $-$100 & $-$68 \\
004500.7+412836.9 & 4.1 & 3.2 & 3.3 & 3.4 & $-$300 & $-$191 & 
004500.9+413101.1 & 3.0 & 2.8 & 2.8 & 2.9 & $-$300 & \nodata \\  
004503.0+413249.4 & 2.0 & 2.3 & 2.1 & 1.9 & $-$300 & $-$153 & 
004504.6+413237.6 & 1.8 & 2.1 & 2.4 & 2.0 & $-$300 & $-$151 \\
004505.3+413845.9 & 2.3 & 2.5 & 2.4 & 2.7 & $-$100 & \nodata &   
004505.9+413925.5 & 2.2 & 2.4 & 2.3 & 2.3 & $-$100 & $-$126 \\
004506.1+413615.0 & 2.1 & 1.9 & 1.9 & 1.8 & $-$100 & $-$154 & 
004506.1+415121.0 & 3.4 & 2.7 & 3.8 & 2.6 & $-$100 & $-$52 \\
004506.2+413424.4 & 3.0 & 2.6 & 2.6 & 2.7 & $-$100 & $-$150 & 
004506.9+413407.8 & 2.8 & 2.7 & 2.7 & 2.7 & $-$100 & $-$140 \\ 
004507.5+413439.4 & 2.1 & 2.4 & 2.4 & 2.4 & $-$100 & $-$144 & 
004508.2+413424.2 & 2.2 & 2.5 & 2.3 & 2.4 & $-$100 & $-$132 \\
004509.0+415209.7 & 3.3 & 2.7 & 2.6 & 2.7 & $-$100 & $-$62 & 
004510.0+420143.6 & 2.3 & 2.7 & 2.4 & 2.5 & $-$100 & \nodata \\  
004510.3+420228.5 & 2.6 & 2.5 & 2.5 & 2.6 & $-$100 & \nodata & 
004510.4+413716.0 & 1.9 & 2.2 & 2.2 & 2.1 & $-$100 & $-$104 \\ 
004510.5+413426.7 & 2.4 & 2.4 & 2.2 & 2.5 & $-$100 & $-$121 & 
004510.8+415938.9 & 2.3 & 2.3 & 2.2 & 2.1 & $-$100 & \nodata \\
004511.2+413644.9 & 3.3 & 2.7 & 2.5 & 2.6 & $-$100 & $-$116 & 
004511.3+413633.9 & 2.7 & 2.7 & 2.9 & 2.6 & $-$100 & $-$136 \\
004511.6+420130.3 & 2.0 & 2.1 & 2.4 & 2.1 & $-$100 & \nodata &   
004512.1+415542.5 & 1.9 & 2.0 & 2.2 & 2.0 & $-$100 & $-$55 \\
004512.4+413709.6 & 3.2 & 2.7 & 2.8 & 2.6 & $-$100 & $-$129 & 
004512.8+413531.6 & 2.4 & 1.9 & 2.1 & 2.0 & $-$100 & $-$133 \\ 
004514.4+413723.6 & 2.5 & 2.4 & 2.3 & 2.6 & $-$100 & $-$120 & 
004515.2+413948.5 & 3.1 & 2.5 & 2.6 & 2.5 & $-$100 & \nodata \\
004515.9+420254.4\tablenotemark{a} & 2.3 & 2.4 & 2.3 & 2.3 & $-$100 & \nodata & 
004518.1+413920.5 & 2.2 & 2.5 & 2.6 & 2.2 & $-$100 & $-$86 \\
004518.5+414013.2 & 2.1 & 2.1 & 2.1 & 1.8 & $-$100 & \nodata & 
004518.7+413906.1 & 2.2 & 2.1 & 2.4 & 2.2 & $-$100 & \nodata \\  
004518.8+420331.8 & 1.2 & 1.4 & 1.3 & 1.5 & $-$100 & \nodata &   
004520.7+414716.7 & 3.1 & 2.5 & 2.8 & 2.5 & $-$100 & $-$100 \\
004520.9+414248.8 & 2.3 & 2.2 & 2.3 & 1.9 & $-$100 & \nodata &   
004521.6+420345.1 & 2.3 & 2.3 & 2.4 & 2.3 & $-$100 & \nodata \\  
004523.1+414346.0 & 2.0 & 2.0 & 2.1 & 2.3 & $-$100 & $-$103 & 
004524.4+415537.4 & 2.1 & 1.8 & 2.1 & 3.6 & $-$100 & \nodata \\  
004526.8+415820.1 & 2.6 & 2.0 & 2.1 & 2.6 & $-$100 & \nodata & 
004527.0+415135.5 & 1.8 & 2.4 & 2.2 & 2.0 & $-$100 & \nodata \\
004528.0+415928.1 & 2.6 & 2.6 & 2.3 & 2.3 & $-$100 & $-$50 & 
004528.2+414513.6 & 3.1 & 2.8 & 2.4 & 2.8 & $-$100 & $-$91 \\
004528.2+414630.6 & 1.7 & 2.1 & 2.3 & 2.1 & $-$100 & $-$75 & 
004528.6+415000.2 & 2.1 & 2.0 & 2.3 & 2.0 & $-$100 & $-$55 \\
004532.2+414543.3 & 2.0 & 2.5 & 2.2 & 1.9 & $-$100 & $-$94 & 
004533.3+414739.8 & 2.0 & 2.4 & 2.3 & 2.1 & $-$100 & $-$61 \\
004533.6+414728.2 & 2.3 & 2.3 & 2.4 & 2.5 & $-$100 & $-$70 & 
004534.1+414703.3 & 2.3 & 2.5 & 2.2 & 2.5 & $-$100 & \nodata \\  
004536.3+414252.0 & 2.1 & 2.3 & 2.3 & 2.4 & $-$100 & $-$129 & 
004536.5+415307.7 & 2.2 & 2.0 & 2.2 & 2.0 & $-$100 & $-$78 \\
004536.9+415704.0 & 3.2 & 2.7 & 2.7 & 2.4 & $-$100 & $-$72 & 
004537.2+415802.4 & 2.9 & 2.6 & 2.8 & 3.0 & $-$100 & $-$36 \\
004537.3+415107.0 & 3.2 & 2.9 & 2.7 & 2.6 & $-$100 & \nodata &   
004537.6+415424.1 & 3.2 & 2.8 & 3.1 & 2.8 & $-$100 & $-$67 \\
004538.5+415231.1 & 3.2 & 2.9 & 2.9 & 2.4 & $-$100 & $-$45 & 
004540.0+415510.2\tablenotemark{b} & 3.2 & 2.7 & 2.5 & 2.9 & $-$100 & $-$63 \\
004541.4+415550.4 & 1.9 & 2.0 & 2.0 & 2.2 & $-$100 & $-$59 & 
004541.6+415107.7 & 3.3 & 2.5 & 2.6 & 2.6 & $-$100 & $-$89 \\
004542.9+415234.8 & 2.5 & 2.0 & 2.0 & 2.0 & $-$100 & $-$62 & 
004543.3+415109.3 & 1.1 & 1.2 & 1.2 & 1.2 & $-$100 & \nodata \\   
004543.3+415301.1 & 3.0 & 2.7 & 2.6 & 2.7 & $-$100 & $-$57 & 
004543.5+414235.1 & 2.0 & 2.3 & 2.0 & 2.0 & $-$100 & \nodata \\
004544.3+415207.4 & 3.1 & 2.9 & 2.8 & 2.8 & $-$100 & $-$31 & 
004549.7+421017.1 & 2.3 & 2.3 & 2.4 & 2.0 & $-$100 & \nodata \\
004555.2+415645.8 & 2.1 & 2.2 & 2.5 & 2.1 & $-$100 & \nodata & 
004608.5+421131.0 & 2.9 & 2.6 & 3.0 & 2.7 & $-$100 & \nodata \\
004613.4+415224.4 & 2.1 & 2.5 & 2.4 & 2.2 & $-$100 & \nodata &   
004617.6+415158.0 & 3.4 & 2.7 & 2.7 & 2.8 & $-$100 & $-$122 \\
004623.9+421215.2 & 2.1 & 2.4 & 2.5 & 2.5 & $-$100 & \nodata & 
004625.4+421156.0 & 2.2 & 2.3 & 2.3 & 2.3 & $-$100 & \nodata \\
004626.0+421121.7 & 2.2 & 2.3 & 2.2 & 2.3 & $-$100 & \nodata & 
004627.9+415920.4 & 2.2 & 2.4 & 2.3 & 2.2 & $-$100 & \nodata \\
004631.5+421342.6 & 2.4 & 2.5 & 2.4 & 2.3 & $-$100 & \nodata & 
004633.4+421244.2 & 2.5 & 2.3 & 2.4 & 2.2 & $-$100 & \nodata \\
004633.6+415932.0 & 3.0 & 2.8 & 2.8 & 2.8 & $-$100 & \nodata & 
004634.2+415636.8 & 1.9 & 2.1 & 2.2 & 2.1 & $-$100 & \nodata \\
004634.4+421143.1 & 2.4 & 2.0 & 2.0 & 1.9 & $-$100 & \nodata & 
004641.6+421156.2\tablenotemark{b} & 3.2 & 2.7 & 2.7 & 2.6 & $-$100 & \nodata \\
004641.9+421547.8 & 2.3 & 2.4 & 2.5 & 2.4 & $-$100 & \nodata & 
004642.2+415837.3 & 2.4 & 2.5 & 2.3 & 2.3 & $-$100 & \nodata \\
004642.6+421406.8\tablenotemark{a} & 2.4 & 2.5 & 2.5 & 2.6 & $-$100 & \nodata & 
004645.9+420453.1 & 2.2 & 2.2 & 2.3 & 2.3 & $-$100 & \nodata \\
004654.5+420046.2 & 2.3 & 2.2 & 2.6 & 2.2 & $-$100 & \nodata & 
004703.1+415755.4\tablenotemark{a} & 2.3 & 2.3 & 2.5 & 2.3 & $-$100 & \nodata 
\enddata
\tablenotetext{a}{Giant star identified by \citet{amiri16}.}
\tablenotetext{b}{Planetary nebula identified by \citet{merrett06}.}
\tablecomments{Detected H$_2$O maser regions are in bold.  
The rms noise is quoted per 3.1--3.3~km~s$^{-1}$ (244 kHz) channels.
$V_{\rm Obs}$ is the observed central velocity tuning of the 632--674~\kms\ (50~MHz) bandwidth.
$V_{\text{CO}}$ uncertainties are typically $\sim$1 km s$^{-1}$ \citep{nieten06}, 
but the CO velocity map shows mean and median dispersions of 19 and 13~km s$^{-1}$ within the GBT beam.  
All velocities are in the heliocentric reference frame using the optical definition.  
}
\end{deluxetable*}

Observations were conducted with the dual K-band receivers in 2010 and beams 
1 and 2 of the K-band focal plane array in 2011--2012 in a
nodding mode in two circular polarizations with 0.15--0.16~km~s$^{-1}$ (12.2~kHz)
channels and 9-level sampling.    The time on-source was 5 minutes 
except for sources that were re-observed to confirm or refute
possible lines (typically 10 minutes).
A winking calibration diode and hourly atmospheric opacity estimates were used for flux
density calibration.  Opacities ranged from 0.03 to 0.14 nepers but
were typically 0.06 nepers.  The estimated uncertainty in the flux
density calibration is $\sim$20$\%$.
Pointing and focus corrections were made hourly.
Pointing was typically good to within a few arcseconds, and the
largest pointing corrections during observing sessions were no more
than 6\arcsec.  
The resolution of the 24~\mum\/ Spitzer image is 6\arcsec\/ \citep{gordon08}, so the unresolved 
IR sources remained within the 33\arcsec\ GBT beam even during the largest pointing drifts.
The 33\arcsec\ beam (FWHM) at 22~GHz spans 125~pc in M31.

After averaging polarizations, 
spectra were Hanning smoothed and subsequently Gaussian smoothed to obtain a final
spectral resolution of 3.1--3.3~\kms\ (244~kHz) for the line search.  H66$\alpha$ was
also smoothed to 9.8~km~s$^{-1}$ (732~kHz) to search for broad recombination lines
\citep[e.g.,][]{dieter67,mezger67}.  
Polynomial baselines, typically of fifth order, were fit and subtracted
to obtain flat and generally uniform-noise spectra.
Spectral rms noise measurements for each pointing center and observed line are listed in Table
\ref{tab:obs}.
Non-detection spectra generally did not show any features greater than
3$\sigma$, corresponding to roughly 10~mJy, at least an order of magnitude
more sensitive than previous surveys \citep[][Claussen \& Beasley,
priv.\ comm.]{greenhill95,imai01}, and likely the weakest line that can 
currently be used for VLBI proper motion studies.
All data reduction was
performed in GBTIDL\footnote{GBTIDL (\url{http://gbtidl.nrao.edu/}) is 
the data reduction package produced by NRAO 
and written in the IDL language for the reduction of GBT data.}.

\begin{figure}
\includegraphics[width=0.98\textwidth,trim=0.33in 1.95in -0.6in 1.2in,clip]{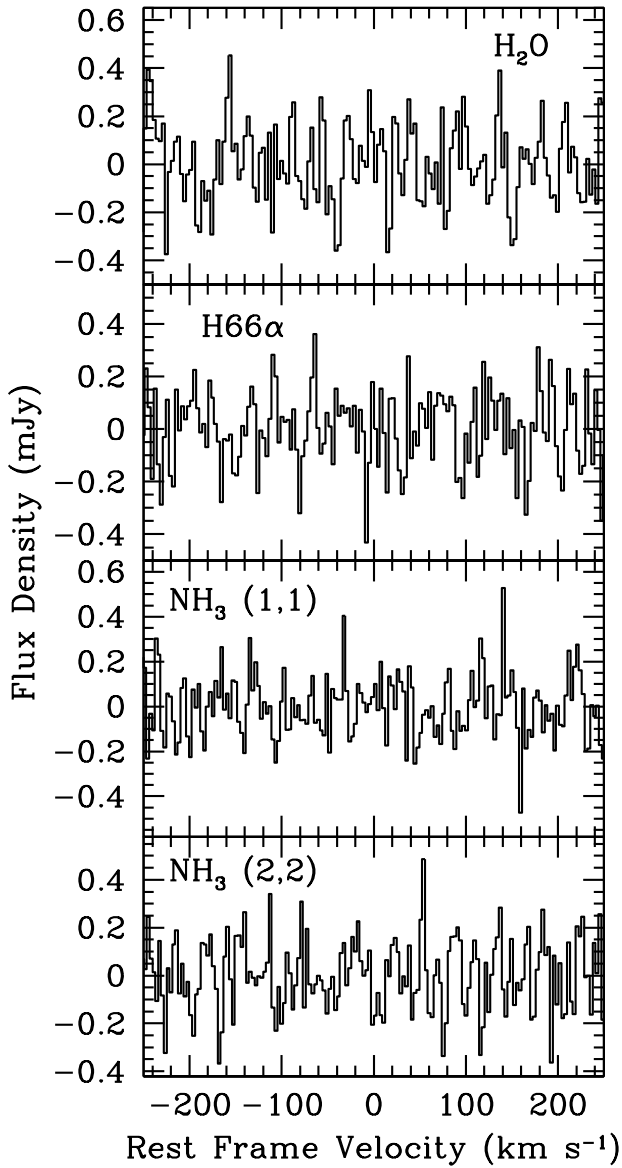}
\caption{Mean stacks of 300 spectra (299 spectra for H$_2$O).  
Velocities of individual spectra were centered on the local CO velocity measured by \citet{nieten06} prior
to stacking.  Channels are 3.1--3.3~km~s$^{-1}$.  Median stacks show a similar lack of significant lines and
12--21\% higher noise.  
\label{fig:stacks}}
\end{figure}

In order to assess the average and median line emission of the sample, 
we aligned H$_2$O, NH$_3$, and H66$\alpha$ spectra 
according to the \citet{nieten06} CO velocities and median- and mean-stacked 
them to determine whether the median or mean object selected for the survey 
produces line emission in a given species an order of magnitude below the single-spectrum 
sensitivity.  Spectra were noise-weighted in the mean stacks, and 300 spectra were used 
for both the mean and median stacks (except for the H$_2$O stacks, which used 299 spectra).  
Omitted spectra either had no CO velocity information, had unreliable CO velocity information, 
or have been identified to be other than an \ion{H}{2} region (indicated with footnotes in Table \ref{tab:obs}; see 
also Section \ref{sec:results}).  
Every object in Table \ref{tab:obs} was manually inspected in the \citet{nieten06} CO 
velocity map to ensure that no mask edges, incomplete data, or high (instrumental) 
velocity dispersion regions were included in the stack.  
All objects in Table \ref{tab:obs} that show a CO velocity can be considered reliable.  

\section{Results}\label{sec:results}

We have detected water maser emission in five out of 506
24~\mum-selected regions in M31 (Table \ref{tab:obs}).  
The water maser isotropic line luminosity sensitivity is
$4.4\times10^{-4}$~$L_\odot$, which is a 3$\sigma$ peak line flux density limit of 9~mJy and width 3.3~\kms.
The detected water masers were presented in \citet{darling11}
and are not reproduced here.  Omitting the 44 objects 
likely to be planetary nebulae \citep[9 objects;][]{merrett06,amiri16} or giant stars \citep[35 objects;][]{amiri16}, 
the overall detection rate is 1.1(0.5)\%.  
The actual water maser 
fraction likely depends on star formation stage and conditions, and 
our overall detection rate is a strong function of the survey sensitivity; 
\citet{amiri16} demonstrate that we detect only the high luminosity tail of
the underlying maser population.   Conversely, it is unlikely
that all of the remaining 24~\mum\ sources are compact \ion{H}{2}
regions \citep[e.g.,][]{verley07,mould08}, 
so the maser fraction among star-forming regions may be higher than 1.1(0.5)\% 
given the survey sensitivity.  
The maser detection rate would also be higher 
were we to use a more refined selection method such as the one presented in \citet{amiri16}.  

No NH$_3$ or H66$\alpha$ lines were detected in the survey; see
Table \ref{tab:obs} for rms noise values.  
Figure \ref{fig:stacks} shows the noise-weighted mean spectral stacks for the four observed spectral 
lines:  H$_2$O, H66$\alpha$, NH$_3$ (1,1), and NH$_3$ (2,2).  
The stacks of 300 spectra (299 spectra for H$_2$O) reached rms noise values of 0.17, 0.14, 0.15, and
0.15 mJy per 3.1--3.3~km~s$^{-1}$ channel, respectively.
The median stacks appear similar to the mean stacks, but with 12, 21, 20, and 13\% higher noise.   No lines
are detected in these stacks, which show the expected $\sqrt{N}$ noise improvement
over single-object spectra.

\section{Discussion}\label{sec:discussion}

Neither the mean nor the median compact 24~$\mu$m-emitting region 
selected for the water maser survey shows detectable emission in the four observed lines.  
In all of the stacked spectra, errors in the CO velocity used to center spectra, CO velocity 
variance, and CO line peak offsets from the denser molecular gas velocity, can all contribute
to incorrect spectral shifts in the stacking process.  These factors will dilute the detectability 
of lines in spectral stacks, particularly in the mean.  

The water maser fraction in Galactic surveys can be of order 50\% \citep[e.g.,][]{kurtz05,urquhart11}, 
so one might expect to detect line emission from the median object in an adequately 
deep survey.  
While the line luminosity sensitivity of the M31 survey does not reach 
the sensitivity needed in Galactic surveys to detect the median water maser in 
individual spectra, the median spectral stack reaches a 3$\sigma$ luminosity sensitivity 
of $2.8\times10^{-5}$~$L_\odot$ (assuming a 3.3~km~s$^{-1}$-wide maser).  Since the 
median \citet{urquhart11} H$_2$O maser isotropic line luminosity is $1\times10^{-5}$~$L_\odot$ 
and since the CO velocity alignment of the stacks lacks the velocity precision needed for 
alignment of few~km~s$^{-1}$ lines, the M31 spectral stack cannot quite detect the
median Galactic \ion{H}{2} water maser equivalent.   

For the H66$\alpha$ line, the non-detection in stacked spectra probably implies that the 
mean or median region simply does not emit in this line.  
Aside from selecting for 22 GHz radio
continuum backlights, the 24 $\mu$m selection method is sub-optimal for hydrogen radio recombination 
lines (the H66$\alpha$ line was included in the GBT observations to allow serendipitous detection).  

NH$_3$ emission, in contrast, is common in \ion{H}{2} regions; \citet{urquhart11} obtain a 
detection rate of $\sim$80\%, fairly independent of \ion{H}{2} region luminosity, and \citet{dunham11} 
detected NH$_3$ (1,1) toward 72\%   of 1.1 mm continuum sources in the inner Galaxy.  
One would expect to detect NH$_3$ lines in mean or median spectra of adequate depth, provided 
the NH$_3$ filling factor does not become too small at the distance of M31.  We discuss 
these issues below.

\subsection{Water Maser Detection Rate and Star Formation}

The sensitivity of the GBT survey corresponds to a $\sim$200~Jy Galactic maser
6~kpc distant.  There are 23 H$_2$O masers in the \citet{palagi93} sample and
roughly 30 masers in the \citet{breen10} sample that our survey could
detect at the distance of M31.  Hence, sensitivity is not the only
consideration; the number of water masers detectable
in M31 will also depend on the 
the substantially lower star formation rate of M31 compared to Galactic.
The maser detection rate, however, should be comparable between galaxies
in similarly-selected samples of star-forming regions if the physical
conditions of star formation are similar.  

In the \citet{urquhart11} Galactic water maser survey, there are 30 H$_2$O masers with peak flux density greater than 200 Jy, and
there are 22 H$_2$O masers with isotropic luminosity greater than $4.4\times10^{-4}$~$L_\odot$. 
Were this Galactic survey restricted to the luminosity sensitivity of the M31 water maser survey, the 
net detection rate would fall from 52(2)\% to 3.7(0.8)\%, which is not significantly different from our
M31 detection rate of 1.1(0.5)\%.  The \citet{urquhart11} 
detection rate, however, is a strong function of \ion{H}{2} region bolometric luminosity, 
approaching 100\% in the highest luminosity regions.  Owing to the order-of-magnitude reduced
star formation rate in M31 compared to the Galaxy, this high luminosity tail of star-forming regions
may not exist in M31, further reducing the expected maser detection rate.

To estimate the total number of expected water masers in Local Group
galaxies, \citet{brunthaler06} used the \citet{greenhill90} Galactic 
water maser luminosity function (LF) scaled from the Galactic star 
formation rate to that of Local Group galaxies, assuming
that maser population is proportional to the star formation rate.  
Local Group water maser
statistics agree fairly well with this approach \citep{brunthaler06}.
Updating \citet{darling11}, we scale the Galactic star formation rate
of $\sim$2~$M_\odot$~yr$^{-1}$ \citep{chomiuk11}
to that of M31, $0.25^{+0.06}_{-0.04}$~$M_\odot$~yr$^{-1}$ \citep{ford13}, 
to predict 3.3 water masers for an isotropic line luminosity limit of 
$4.4\times10^{-4}$~$L_\odot$.
This prediction suffers from small number statistics, but 
our five detections suggest that either we have detected 
the majority of bright water masers in M31 (the high-luminosity tail 
of the water maser LF) or that the star formation 
scaling approach is inaccurate.  The former seems more likely:  \citet{amiri16} show
that we have surveyed nearly all of the regions most likely to produce luminous water masers
and that there are probably no more bright water masers to be found in M31.

\subsection{Ammonia}\label{subsec:ammonia}

While ammonia lines lack the intrinsic 
brightness temperature of H$_2$O, they can be beam-filling.  
Given the mean $T_{\rm mb}=1.1$~K (1,1) line peak and mean distance of $\sim$5~kpc of
the \citet{urquhart11} sample, and using a GBT main-beam efficiency of
0.87 and gain of 1.88 K~Jy$^{-1}$ \citep{mangum13},
the average Galactic NH$_3$ emission from an \ion{H}{2}
region\footnote{Not all NH$_3$-emitting objects in the 
  \citet{urquhart11} sample are \ion{H}{2} regions.}
scaled to the distance of M31 would produce a $\sim$0.02 mJy line peak, 
all other factors being equal and assuming no confusion in the NH$_3$ 
emission from multiple \ion{H}{2} regions along Galactic sight lines
(the \citet{dunham11} measurements provide a 
similar prediction).   There would thus need to be roughly
150 such NH$_3$ emitters within a GBT beam to approach the rms noise
in a single pointing.  The rough size of a Galactic NH$_3$ source in
the \citet{urquhart11} sample is 0.2-0.4 pc, so by area, it is
possible that 150 such objects could fall within the 121 pc GBT beam
in the molecular ring of M31.  The brightest \ion{H}{2} region in the 
\citet{urquhart11} sample would appear at 0.13 mJy in M31, still a
factor of $\sim$20 below the rms noise in individual spectra.  The
NH$_3$ (2,2) line is typically weaker than the (1,1) line, so our
observations of the (2,2) line are less constraining than the (1,1) line. 

The mean-stacked NH$_3$ (1,1) spectra show an rms noise per channel seven times 
higher than the expected average Galactic NH$_3$ (1,1) line emitter, 
but if the average 121 pc-diameter region contained roughly 20 such
regions, then a $3\,\sigma$ peak would be detected.  The data do not
support this.  

Extragalactic NH$_3$ cm metastable lines have been detected toward numerous galaxies, some down
to the few mJy/mK level by, e.g.,
\citet{martin79,takano00,henkel00,weiss01,takano02,mauersberger03,ott05,ott11},
and \citet{mangum13}, but in most cases the galaxies are actively
star-forming --- if not starbursts --- at substantially higher rates than
M31.  The non-detection of NH$_3$ toward M31 is not yet constraining
on the NH$_3$ abundance compared to the Galactic value.

\section{Conclusions}

We have expanded a survey for H$_2$O masers in M31
based on pointed observations of 24~\mum-selected regions.  
While the 24~\mum\/ selection method clearly succeeds in identifying 
water masers (albeit at the low rate of 1.1(0.5)\%), 24~\mum\/ emission
is a necessary but not a sufficient condition for water maser activity
(see \citet{amiri16} for a study of the properties of star-forming 
regions most likely to produce water masers).  
In \citet{darling11}, we suggested that the catalog of water masers
in M31 is incomplete and that an exhaustive survey of IR-luminous regions
would identify additional masers.  The prediction was too optimistic, and it should
be noted that the detected masers are consistent with the aggregate star formation rate 
of M31 given its distance and the sensitivity of the survey.   New masers could still be 
detected in deeper surveys or in future surveys of similar depth, given the variable nature
of water masers in star-forming regions.  However, variability also implies that the currently detected 
masers could themselves be flaring and therefore fade, limiting their astrometric utility.

For now, the five known water masers in M31 will have to serve as the basis for proper motion studies, 
which will probably minimally constrain the three expected proper motion signals 
(plus random peculiar maser motions) described by \citet{darling11}:  
(1) the systemic proper motion \citep[already constrained by][]{sohn12,vandermarel12b}, 
(2) proper rotation, and
(3) proper expansion caused by the $-300$~km~s$^{-1}$ (heliocentric) approach of M31 \citep[see also][]{darling13}.  
 We predict that proper rotation and systemic proper
motion will be measurable in a few years with current facilities, and
that the ``moving cluster'' divergence of maser spots as M31
approaches the Galaxy may be detectable in a decade \citep{darling13}, but 
a possible previous interaction between M31 and M32 may have produced peculiar radial motions
in the molecular ring of M31 that would mask this signal \citep{block06,dierickx14}.

\acknowledgements
All authors acknowledge support from the NSF grant AST-1109078.
The authors thank K.\ Gordon for the Spitzer map, M.\ Claussen and T.\ Beasley for sharing their results, 
and the anonymous referee for helpful comments.
This research has made use of NASA's Astrophysics Data System Bibliographic Services
and the NASA/IPAC Extragalactic Database
(NED) and uses observations made with the {\it Spitzer Space Telescope},
both of which are operated by the Jet Propulsion Laboratory,
California Institute of Technology, under a contract with
NASA. 

\facility{GBT}

\end{document}